\documentclass[pre,twocolumn,aps,showpacs]{revtex4}
\usepackage{dcolumn}
\usepackage{graphicx}
\usepackage{graphicx,amssymb,amsmath}
\usepackage{natbib}
\usepackage{bm}

\begin{document}

\newcommand{\unit}{\bf\hat}
\newcommand{\shear}{\mu}

\title{The mechanical response of semiflexible networks to localized perturbations}

\author{D.A.~Head$^{1,2}$, A.J.~Levine$^{3,4}$ and
F.C.~MacKintosh$^{1}$}

\affiliation{$^{1}$Department of Physics and Astronomy,
Vrije Universiteit, Amsterdam, The Netherlands.\\
$^{2}$Department of
Applied Physics, The University of Tokyo, Tokyo 113-8656, Japan.\\
$^{3}$Department of Physics, University of Massachusetts, Amherst
MA 01060, USA.
$^{4}$Department of Chemistry \& Biochemistry, University of California, Los Angeles, CA 90095}

\date{\today}

\begin{abstract}
Previous research on semiflexible polymers including cytoskeletal
networks in cells has suggested the existence of distinct regimes
of elastic response, in which the strain field is either uniform
(affine) or non--uniform (non--affine) under external stress.
Associated with these regimes, it has been further suggested that
a new fundamental length scale emerges, which characterizes the
scale for the crossover from non--affine to affine deformations.
Here, we extend these studies by probing the response to localized forces and force dipoles.
We show that the previously identified nonaffinity length [D.A. Head {\em et al.}
PRE {\bf 68}, 061907 (2003).] controls the mesoscopic response to point forces and the
crossover to continuum elastic behavior at large distances.
\end{abstract}


\maketitle

\section{Introduction}
\label{s:intro} Semiflexible polymers such as filamentous proteins
resemble elastic rods on a molecular scale, while exhibiting
significant thermal fluctuations on the scale of micrometers or
even less. This has made them useful as model systems allowing for
direct visualization via optical microscopy. But, semiflexible
polymers are not just large versions of their more well-studied
flexible cousins such as polystyrene. Filamentous proteins, in
particular, have been shown to exhibit \emph{qualitatively}
different behavior in their networks and solutions. A fundamental
reason for this is the fact that the thermal \emph{persistence
length}, a measure of filament stiffness as the length at which
thermal bending fluctuations become apparent, can become large
compared with other important length scales such as the spacing
between polymers in solutions, or the distance between chemical
crosslinks in a network.

One of the most studied semiflexible polymers in recent years has
been F-actin, a filamentous protein that plays a key structural
role in cells \cite{alberts:88,pollard:86,elson:88}. These occur
in combination with a wide range of specific proteins for
crosslinking, bundling and force generation in cells. These
composites, together with other filamentous proteins such as
microtubules, constitute the so-called \emph{cytoskeleton} that
gives cells both mechanical integrity and structure.  This
biopolymer gel is but one example of a large class of polymeric
materials that can store elastic energy in a combination of
bending and extensional deformations of the constituent elements.
Such systems can be called semiflexible gels or networks.

One of the important lessons from recent experimental and
theoretical studies is that the shear modulus of cross-linked
semiflexible networks bears a much more complex relationship to
the mechanical properties of the constituent filaments and to the
microstructure of the gel than is the case for flexible polymer
gels \cite{rubenstein:03}.
Recently, it has been shown that semiflexible gels exhibit a striking cross-over
\cite{head:03a,wilhelm:03,head:03b,head:03c,levine:04} between a
response to external shear stress that is characterized by a
spatially heterogeneous strain (a \emph{non-affine} regime
\cite{kroy:96,satcher:96}) and a uniform strain response (an
\emph{affine} regime \cite{mackintosh:95}). This crossover is
governed primarily by cross-link density and molecular weight
(filament length). The bulk shear modulus of the network
simultaneously increases by about six orders of magnitude at this
nonaffine-to-affine cross-over.  The underlying mechanism
responsible for this abrupt cross-over appears to be the
introduction of a new, mesoscopic length scale in the problem that
is related to both the bending stiffness of the constituent
polymers and the mean spacing between consecutive cross-links
along the chain \cite{head:03a,head:03c}.

One can associate this mesoscopic length with the length below
which the deformation of the network departs from the standard
affinity. The nonaffinity length $\lambda$, introduced in Refs.\
\cite{head:03a,head:03c}, can be qualitatively understood as the
typical length over which one finds nonaffine deformation in the
network. In this previous work we presented a scaling analysis
that relates this mesoscopic length scale to the network density
and the stretching and bending moduli of the constituent
filaments.  The macroscopic shear response of the network is then
controlled by a competition between $\lambda$ (the nonaffinity
length) and the filament length, $L$. On the one hand, when the
filament length is long, nonaffine corrections to the deformation
field, which are localized to regions within $\lambda$ of the
filament ends, do not significantly affect the mechanical
properties of the network; the shear modulus of the macroscopic
system is well-described by calculations based on affine
deformation reflecting the fact that nonaffine deformations of the
filament ends are subdominant corrections in this limit. Moreover,
the elastic energy is stored primarily in the (homogeneous)
extension and compression of filaments. On the other hand, when
the filaments are of a length comparable to, or shorter than the
nonaffinity length, {\em i.e.}
$L\stackrel{<}{\scriptstyle\sim}\lambda$ then the nonaffine
deformations of the ends play a large, even dominant role in
determining the mechanical properties of the network. The network
is found to be generally more compliant, and the elastic energy
under applied shear stress is stored in a spatially heterogeneous
manner in the bending of filaments. The existence of these
distinct regimes as a function of filament length reflects a
fundamental difference of these semiflexible polymer networks with
respect to their flexible counterparts: polymers can maintain
their mechanical integrity and state of stress/strain across
network nodes or crosslinks.

These results naturally lead one to pose a number of basic
questions regarding the elastic properties of semiflexible
networks. While these random networks must on basic theoretical
grounds appear as continuum, isotropic materials at the longest
length scales, these considerations do not predict the length at
which the continuum approximation applies.
The previous
identification of the nonaffinity length, $\lambda$  as the
only mesoscopic length associated with the nonaffine-to-affine
cross-over in uniformly sheared semiflexible networks
suggests that this length should more generally control the
cross-over to continuum
behavior \cite{head:03c}. After all the affine deformation of the
network under uniform stress at scales large compared to $\lambda$
shows that in one case at least the nonaffinity length controls
the cross-over to continuum behavior.  One of the principal
results of the present work is the demonstration that $\lambda$ more
generally controls this cross-over to continuum mechanics in
semiflexible gel systems.

Prior work has focussed exclusively on simple shear and uniaxial
extension. In order to better examine the universality of the
previous results, we study the opposite limit of a highly
localized external force in the form of a point force monopole or
dipole. If one can show that the elastic (displacement) Green's
function of the system similarly depends on only one additional
parameter $\lambda$ then it would appear that this quantity
completes the coarse-grained elastic description of the system on
all length scales down to the mean distance between cross-links.
It may be, however, that the deformation field of these
semiflexible networks is much more complex and the simplification
introduced by $\lambda$ in the description of the network's
response to uniform shear strain cannot be generalized to
deformations resulting from more general forcing conditions.

While our results below, indeed, show that $\lambda$ does largely
control the crossover to (the \emph{far field}) continuum
elasticity, the observed elastic Green's function is sensitive to
the local structure of the network on length scales below
$\lambda$. Below we discuss how we quantify the structure of the
Green's function and its approach to the form required by
continuum elasticity. The observed elastic Green's function,
however, depends not only on the $\lambda$--dependent Lam\'{e}
coefficients of the material, but also on local properties of the
displacement field immediately surrounding the the point force. In
effect one may imagine that, upon the application of the point
force, the network acts as a type of composite material: within a
distance $\lambda$ of the point force it deforms in a way not well
described by continuum theories, while outside of that zone it
does appear to act like an elastic continuum.
The complete Green's function depends, of course,  on the material properties of both media.
Unfortunately, only one of those media (the outer zone) is well
characterized by the simple continuum theory of an isotropic
elastic solid, so the complete Green's function remains complex
and depends on the detailed network structure within the inner
zone surrounding the applied force.

There is another set of questions that may be addressed via the
study of the network's response to point forces and force dipoles.
Such forces not only probe the material properties of the network
in a manner complimentary to the uniform strains explored earlier,
they also have direct physical implications for microrheology in
F-actin networks and for the dynamics of the cytoskeleton in
response to the activity of nanoscale molecular motors, {\em e.g.}
myosin.  Fluctuation-based microrheology, an application of the
fluctuation-dissipation theorem to the study of rheology via the
statistical analysis of the thermally fluctuating position of
sub-micron tracer particles embedded in the medium, requires one
to understand the elastic Green's function of the medium. Thus
understanding the response of semiflexible networks to localized
forces has direct experimental implications and consequences for
force production in the cytoskeleton.

In the biological context, the semiflexible network making up the
cytoskeleton is generally found in association with molecular
motors that, to a good approximation, generate transient localized
force dipoles in the material. To both understand force generation
in the cell as well as the material properties of these
cytoskeletal networks driven into nonequilibrium steady states by
these molecular motors, one must determine the displacement field
associated with such motor-induced forces.

Notwithstanding our biological motivation for this work, our
findings also bear on the broader problem of elastic modes in
amorphous materials. It has been shown that the vibrational modes
of deep--quenched Lennard--Jones systems approach a continuum
description only on scales exceeding some mesoscopic
length~$\xi$~\cite{Wittmer:2002,Tanguy:2002}; for the protocols
considered, a value $\xi\sim30$ particle dimensions was robustly
found. This was physically identified with a length scale for
non--affinity, suggesting a direct correspondence with
our~$\lambda$ (although our $\lambda$ can be controlled by varying
the mechanical properties of the constituents). A comparable
length was also found to control the self--averaging of the
Green's function to the form expected by continuum
elasticity~\cite{Leonforte:2004}. These findings for radially
interacting particles are broadly in keeping with our own
investigations for semiflexible polymer networks.
We also mention here that the relationship between continuum elasticity and the 
Green's function has also been discussed for mildly disordered
spring networks~\cite{Goldenberg:2002}.

The remainder of this paper is organized as follows: In section
\ref{s:model} we develop our model of semiflexible, permanently
cross-linked gels, summarize the numerical simulations used to
study it, and discuss the expected structure of the displacement
field when averaged over numerous realizations of the network.  In
section \ref{s:results} we report our results for both point
forces in \ref{ss:point-force} and force dipoles in
\ref{ss:dipole}. We then discuss our studies of the bulk elastic
properties of these networks in \ref{ss:bulk} before concluding in
section \ref{s:discussion}.

\section{Model}
\label{s:model}

\subsection{The semiflexible network}
\label{ss:network}

A highly successful continuum continuum model of individual
semiflexible polymers is the {\em worm--like chain}. This treats
the linear filaments as elastic rods of fixed contour length and
negligible thickness, so that the dominant contribution to the
elastic energy comes from bending modes and the Hamiltonian
linearized to small deviations from a straight configuration that
is given by
\begin{equation}
{\cal H}^{\perp}=\frac{1}{2}\kappa\int (\nabla^{2}u)^{2}\,{\rm
d}s\,, \label{e:H_WLC}
\end{equation}
where $u$ is the transverse displacement of the filament relative
to an arbitrary straight axis, $s$ is its arc length, and the
elastic modulus $\kappa$ gives the bending energy per unit length
$\delta s$.

The longitudinal response of the wormlike chain model is
calculated from the increase in free energy due to an extensional
stress applied along the mean filament axis
\cite{bustamante:94,mackintosh:95}. However, the numerical
algorithm used in our simulations is based on the minimization of
the Hamiltonian of the system, and hence is fundamentally
athermal. The reason for this choice is essentially one of
efficiency: assuming there are no subtle convergence issues,
minimization is expected to be faster than stochastic modelling
and, it is anticipated, give better statistics for a given CPU
time. This does, however, mean that the entropic mechanism
governing the longitudinal response is absent, and an explicit
energetic term is required.

An unconstrained filament at $T=0$ forms a straight configuration,
and thus elongation of its end--to--end distance must be
accompanied by a change in the absolute contour length. It is
therefore natural to incorporate longitudinal modes by adding a
second elastic term to the Hamiltonian for the extension or
shortening of the filament backbone,
\begin{equation}
{\cal H}^{\parallel}=\frac{1}{2}\mu\int \left( \frac{{\rm
d}l(s)}{{\rm d}s} \right)^{2} \,{\rm d}s\,, \label{e:H_Hooke}
\end{equation}
where $dl(s)/ds$ gives the strain or relative change in local
contour length, and $\mu$ is the Young's modulus of the filament
(essentially a spring constant normalized to 1/[length]). Of
course, such modes also exist in thermal systems, but may be
dominated by the entropic spring terms \cite{mackintosh:95},
except possibly for very short filament segments or very densely
cross-linked gels. The connection between thermal/entropic and
athermal longitudinal compliance is discussed in greater detail in
Ref.\ \cite{head:03c}.

The two elastic coefficients $\kappa$ and $\mu$ together define a
length scale $l_{\rm b}=\sqrt{\kappa/\mu}$, which will shall refer
to as the intrinsic {\em bending length} by observing that an
isolated filament constrained to have different tangents at its
end points will deform with this characteristic length. To avoid
potential confusion, however, we note that this is \emph{not} the
typical length scale for bending deformations of a semiflexible
filament. Rather, the bending energy of filaments tends to make
the longest unconstrained wavelength bending mode the dominant
one. Thus, for instance, in a crosslinked gel, filaments are
expected to be bent primarily on a length comparable to the
distance between crosslinks. Nonetheless $l_{\rm b}$ is a useful
measure of filament rigidity, in that large $l_{\rm b}$
corresponds to rigid filaments, and small $l_{\rm b}$ to flexible
ones.

Although $\kappa$ and $\mu$ have been introduced as fundamental
coefficients, if the filament is regarded as a continuous elastic
body with uniform cross section at zero temperature, then they can
both be expressed in terms of the characteristic filament radius
$a$ and intrinsic bulk modulus $Y_{\rm f}$ as $\kappa\sim Y_{\rm
f}a^{4}$ and $\mu\sim Y_{\rm f}a^{2}$. Thus $l_{\rm b}\sim a$, and
thinner filaments are more flexible than thick ones (as measured
by $l_{\rm b}$), as intuitively expected. Given the possibility of
entropic effects in $\mu$, however, we shall treat these as
independent parameters of the theory \cite{head:03c}

The gel is constructed by depositing filaments of monodisperse
length $L$ and zero thickness onto a two--dimensional substrate.
The center--of--mass position vector and orientation of the
filaments are uniformly distributed over the maximum allowed
range, so the system is macroscopically isotropic and homogeneous.
Whenever two filaments overlap they are cross-linked at that
point. Deposition continues until the required mass density, as
measured by the mean distance between crosslinks $l_{\rm c}$, has
been reached. The network thus constructed can be described by
three lengths: $L, l_{\rm b}$, and $l_{\rm c}$ and one modulus
scale: $\mu$. In two-dimensions $l_{\rm c}$ characterizes both the
mass density and cross-link density in spatially random, isotropic
networks. The length $l_{\rm b}$ characterizes the mechanical
properties of the constituent filaments via the ratio of their
bending to stretching compliance. The overall modulus scale $\mu$
will be absorbed into the point forces applied to the network.

Previously \cite{head:03a,head:03c}, we identified an additional
length $\lambda=l_{\rm c}\left(l_{\rm c}/l_{\rm b}\right)^z$,
where $z\simeq 1/3$. This \emph{non--affinity length}
characterized the crossover from non--affine to affine network
response. Specifically, for filament lengths $L$ much larger than
$\lambda$ (\emph{i.e.}, high molecular weight), the bulk network
properties could be understood quantitatively in terms of affine
strains, while significant non--affine effects were observed for
$L\alt\lambda$. Although this length arose naturally from
considerations of bulk network properties \cite{head:03a}, its
dependence on $l_{\rm c}$ and $l_{\rm b}$ can be understood in
terms of a balance of stretching/compression and bending energies
of a single filament, treated in a self--consistent way within a
network \cite{head:03c}. It is important to note that this
(material) length is intermediate, between the (geometric) network
length $l_{\rm c}$ and the macroscopic scale. In fact, in dilute
networks, for which $l_{\rm c}\gg l_{\rm b}$, we shall argue below
that the network can be thought of on scales $\agt\lambda$ as a
quasi--continuum: continuous, as opposed to discrete, but not
necessarily described by macroscopic continuum elasticity. We
shall find that this non--affinity length will play a key role in
our analysis of the displacement field on this
intermediate/quasi--continuum scale.

In order to apply the point forces, a crosslink is chosen at
random and identified as the origin of the system; it is this
crosslink that will later be perturbed. A fixed circular boundary
at radius $R$ from the origin is imposed, and any filaments or
filament segments that extend beyond the boundary are simply
removed or truncated, respectively. Filaments ending on the rigid
boundary are fixed there by another freely rotating bond as are
found at all cross-links in the system. The allowed free rotation
at the boundary means that the boundary supplies arbitrary
constraint forces on the network but cannot support any localized
torques.

\subsection{Numerical method}
\label{ss:numerics}

Details of the simulation method have been presented
elsewhere~\cite{head:03c}. Here we briefly summarize the
procedure, with particular attention on those aspects that are
central to the problems studied in this paper.

The system Hamiltonian ${\cal H}(\{{\bf x_{i}}\})$ is constructed
from discrete versions of (\ref{e:H_WLC}) and (\ref{e:H_Hooke})
applied to the geometry generated by the random deposition
procedure described above. The degrees of freedom $\{{\bf
x_{i}}\}$, or `nodes,' are the position vectors of both crosslinks
and midpoints between crosslinks, the latter so as to incorporate
the first bending mode along the filament segments. Additional
nodes could be included at the cost of additional run time, but
are expected to have a small effect, since the longest
unconstrained wavelengths tend to dominate bending deformations.
Crosslinks are treated as constraints on the relative position of
each connected filament segment, but not on their relative
rotation. Physically this corresponds to an inextensible but
freely rotating linkage. As previously noted, constrained bending
at crosslinks has a small effect except at high network
concentrations (specifically, when $l_{\rm c}$ becomes comparable
to $a$) \cite{head:03c}. Nodes on the boundary are immobile. Note
that, as in our earlier work~\cite{head:03a,head:03b,head:03c},
the network is assumed to be initially unstressed on both
macroscopic and microscopic length scales.

There are two ways in which the system may be perturbed. The
first, which we call {\em monopole} forcing, is to apply an
arbitrarily small external force $\delta{\bf f}^{\rm ext}$ to the
crosslink at the origin. The network is then allowed to relax to a
new configuration consistent with mechanical force balance at
every node. The second, which we call {\em dipole} forcing, is to
introduce a geometrical defect into the system by moving the
central crosslink along the contour length of one of the filaments
to which it belongs, but not the other. Physically, this
corresponds, \emph{e.g.}, to a motor introducing relative motion
of one filament with respect to another filament. In the
simulations, this effect is incorporated by infinitesimally
`shifting' the image of the central node with respect to other
nodes on one filament.

\begin{figure}[htbp]
\centering
\includegraphics[width=8cm]{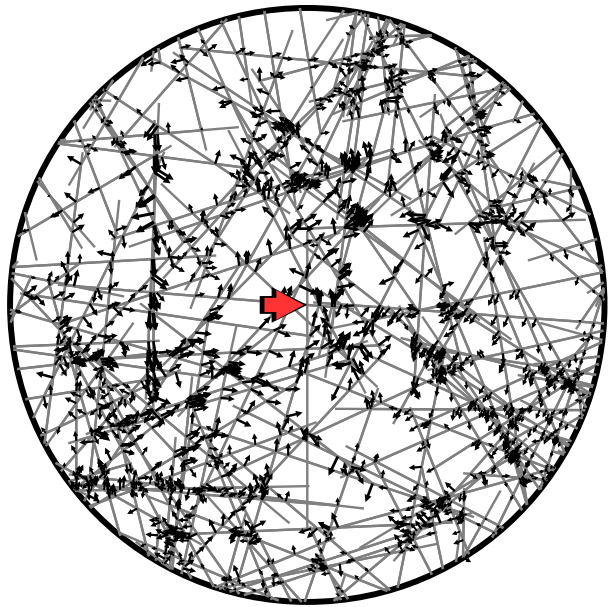}
\caption{{\em (Color online)} An example of a network of filaments
of uniform length $L$ (grey line segments) perturbed by an
external force, denoted by the large red arrow, applied to the
crosslink at the origin of a circular system of radius $R=L$. The
arrow lengths are logarithmically calibrated to the magnitude of
the displacement of at each crosslink. In this example, $L/l_{\rm
c}\approx29.1$, $\lambda/L\approx0.191$ and the force is
perpendicular to one of the filaments that form the central
crosslink; forces can also be applied parallel to a filament. }
\label{f:eg_indiv_run}
\end{figure}

Once the perturbation has been specified, the displacements of the
nodes in the new mechanical equilibrium are calculated by
minimizing the system Hamiltonian ${\cal H}(\{{\bf x_{i}}\})$
using the conjugate gradient method~\cite{NumRec}. This generates a displacement
field for the particular geometry under consideration, as shown in
Fig.~\ref{f:eg_indiv_run}. Note that ${\cal H}(\{{\bf x_{i}}\})$
is linearized about small nodal displacements $\{\delta{\bf
x}_{i}\}$ from their original positions $\{{\bf x}_{i}\}$, so
linear response is assured. The bulk response to non--linear
strains has been recently studied by Onck {\em et al.}~\cite{Onck:2005}.

\subsection{Decomposition of mean displacement field}
\label{ss:decomposition}

\begin{figure}[htpb]
\centering
\includegraphics[width=8cm]{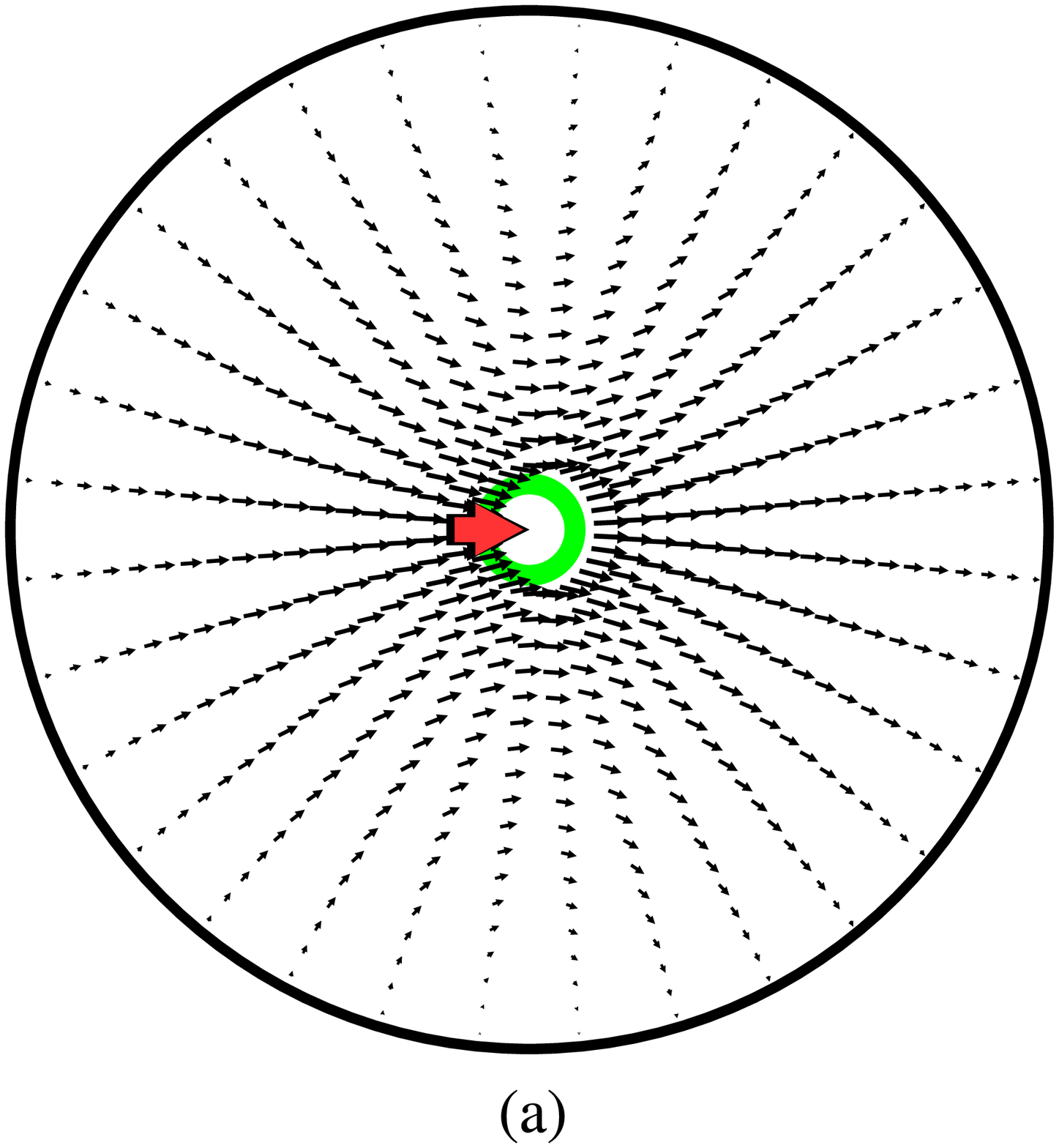}
\includegraphics[width=8cm]{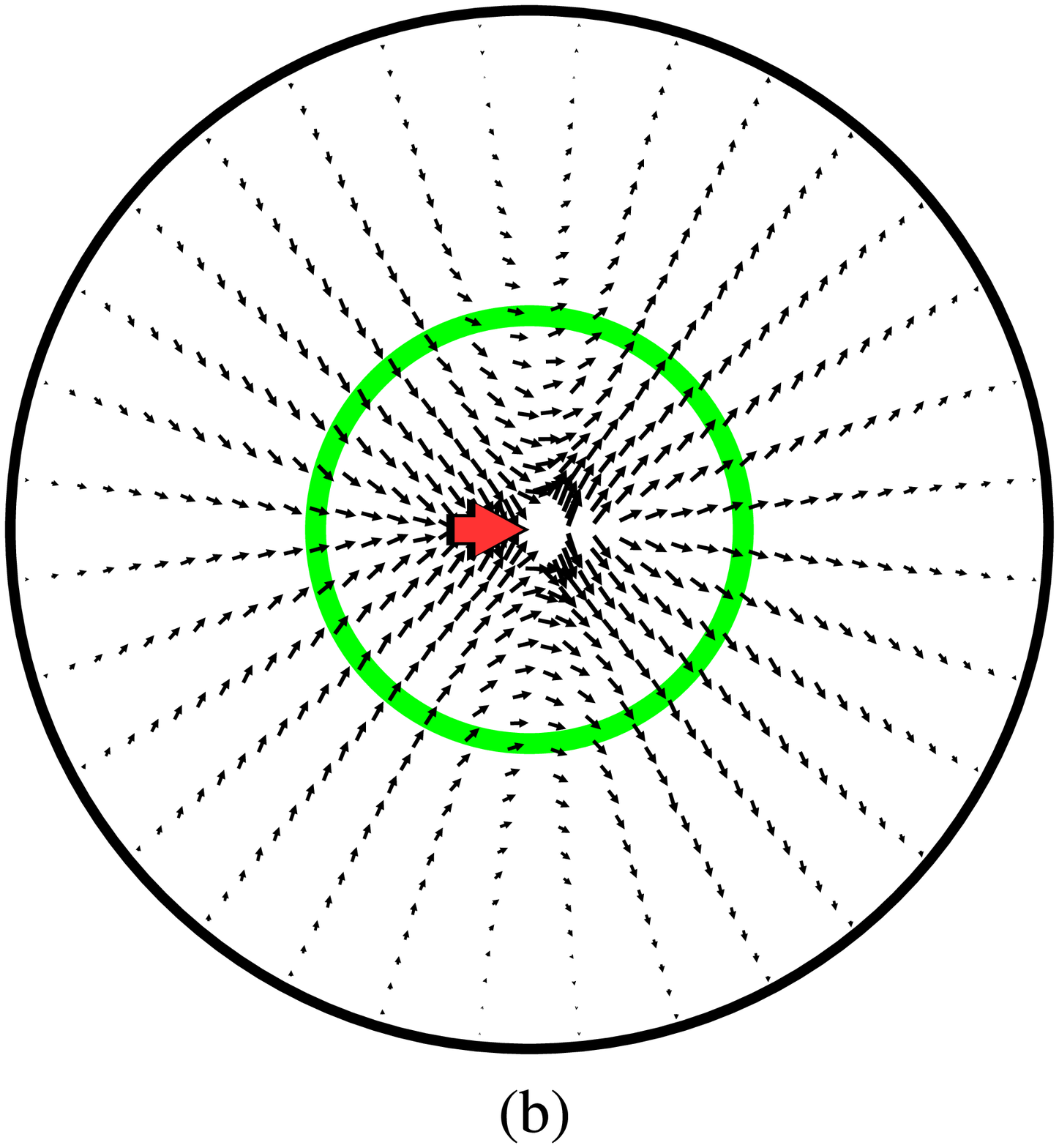}
\caption{{\em (Color online)} Mean response after averaging ${\cal
O}(10^5)$ networks with the same parameters as in
Fig.~(\ref{f:eg_indiv_run}) (including one filament fixed
perpendicular to the external force) except for $\lambda$, which
is {\em (a)}~$\lambda/L\approx0.089$ and {\em
(b)}~$\lambda/L\approx0.42$. For easy visualization, a green
circle of radius $\lambda$ has been inserted into the background
of each plot. Vectors near the center of each system have not been
plotted for clarity. } \label{f:eg_averaged}
\end{figure}

As shown in Fig.~\ref{f:eg_indiv_run}, the displacement field for
a particular network is quite complex and generically shows
anti-correlations between displacements and the local mass
distribution. Although these fluctuations reflect inherent and
possibly interesting physical properties of the gels, a more basic
and immediately applicable quantity to measure is the {\em mean}
displacement field, found by averaging many individual runs with
differing geometries but identical system parameters $R$, $L$,
$l_{\rm c}$ and $\lambda$ (or equivalently $R$, $L$, $l_{\rm c}$
and $l_{\rm b}$). Two examples for differing $\lambda$ (the radius
of the green circle centered on the point of force application)
are given in Fig.~\ref{f:eg_averaged}. These plots demonstrate one
of the primary results of this paper, namely that the crossover
between continuum (or quasi--continuum) response at large lengths,
to a more exotic displacement field at shorter lengths, happens at
a length of order $(\lambda)$ with a prefactor close to unity.

In order to precisely describe the structure of the deformation
field, we consider its most general possible form.  For monopole
forcing, the displacement field ${\bf u}({\bf r})$ at position
vector ${\bf r}$ relative to the origin can be projected onto 3
other vectors, namely the direction of the external force
$\unit{f}$, the unit position vector $\unit{r}$ and the axis
$\unit{n}$ of one of the filaments to which the crosslink is
attached,
\begin{eqnarray}
u_{i}&=&G_{ij}{\hat f}_{j}\,,
\nonumber\\
G_{ij}&=& \frac{f}{\shear_{\rm aff}} \left\{
g^{(r)}\hat{r}_{i}\hat{r}_{j} + g^{(n)}\hat{n}_{i}\hat{n}_{j} +
g^{(f)}\delta_{ij} \right\} \label{e:defn_g}
\end{eqnarray}
where $f$ is the magnitude of the external force, and $\shear_{\rm
aff}$ is the shear modulus as predicted for affine deformation.
This depends on $L$ and $l_{\rm c}$ but {\em not} $\lambda$, and
is included here to factor out the density dependence of the
response. As defined, the $g^{(\cdot)}$ (and the $h^{(\cdot)}$
below) are dimensionless quantities in two dimensions.

Each $g^{(\cdot)}$ can be further decomposed into angular modes in
$\theta$, where $\cos\theta=\unit{n}\cdot\unit{r}$,
\begin{equation}
g^{(\cdot)} = g^{(\cdot)}_{0} + 2\sum_{m>0,\atop m\,{\rm
even}}g^{(\cdot)}_{m}\cos(m\theta). \label{e:g_modes}
\end{equation}
Terms in $\sin(m\theta)$ vanish since the ensemble--averaged
response must be invariant under $\theta\leftrightarrow-\theta$,
and $\cos(m\theta)$ terms with $m$ odd also vanish due to
$\unit{n}\leftrightarrow-\unit{n}$ invariance. This latter
symmetry may appear to violate the known polarity of typical
semiflexible biopolymers such as F-actin and
microtubules~\cite{bray:01}; however, we are interested here in
the {\em mechanical} properties of the filaments, which, within
the approximations of our model, are indeed invariant under
$\unit{n}\leftrightarrow-\unit{n}$.

For dipole forcing, the above procedure is followed with the
additional simplification that ${\unit n}$ is proportional to
${\unit f}$, since the displacement (and hence force) dipole
induced by the motion of a motor will always be parallel to one
filament axis. The decomposition is therefore somewhat simpler,
\begin{eqnarray}
u_{i}&=&H_{ij}{\hat n}_{j}\,,
\nonumber\\
H_{ij}&=& \frac{f}{\shear_{\rm aff}} \left\{
h^{(r)}\hat{r}_{i}\hat{r}_{j} + h^{(n)}\delta_{ij} \right\}\,.
\label{e:h_modes}
\end{eqnarray}
The scalar $f$ is the magnitude of the force dipole; since it is
actually a displacement that is imposed, $f$ is unknown as will be
treated as a fitting parameter. The angular decomposition is
identical to before,
\begin{equation}
h^{(\cdot)} = h^{(\cdot)}_{0} + 2\sum_{m>0,\atop m\,{\rm
even}}h^{(\cdot)}_{m}\cos(m\theta)\,.
\end{equation}

Later sections will refer to the continuum solution for each of
the two forms of forcing. These are given in the appendix. For the
monopole case, only two continuum modes are non--zero, namely
$g^{(r)}_{0}$ and $g^{(f)}_{0}$. We shall refer to these
components of the Green's function as \emph{continuum modes} in
order to distinguish from those components (\emph{non--continuum
modes}) that must vanish in the continuum. We do this even though
for our finite systems the \emph{non--continuum modes} do not
vanish.

The dipole modes are slightly more subtle: after averaging of many
dipole fields generated by the simulation, the resulting field is
{\em quadrupolar}. This is an immediate consequence of the means
of forcing the system. Recall that in an elementary step, a motor
moves parallel to one filament axis. This has the effect of
compressing the filament in front of the dipole, while stretching
the trailing segment. Thus two filament segments are perturbed,
each of which can be treated as a force dipole which, for a
particular network geometry, will be of different magnitudes and
hence the resulting field is dipole. However, the net bias of this
dipole is symmetrically distributed around zero, and thus vanishes
after averaging, leaving a quadrupole field as shown in
Fig.~\ref{f:eg_quadrupole}. As derived in the appendix, the
non--zero modes for this field are $h_{0}^{(r)}$, $h_{2}^{(r)}$,
$h_{0}^{(f)}$ and $h_{2}^{(f)}$.

\begin{figure}[htpb]
\centering
\includegraphics[width=8cm]{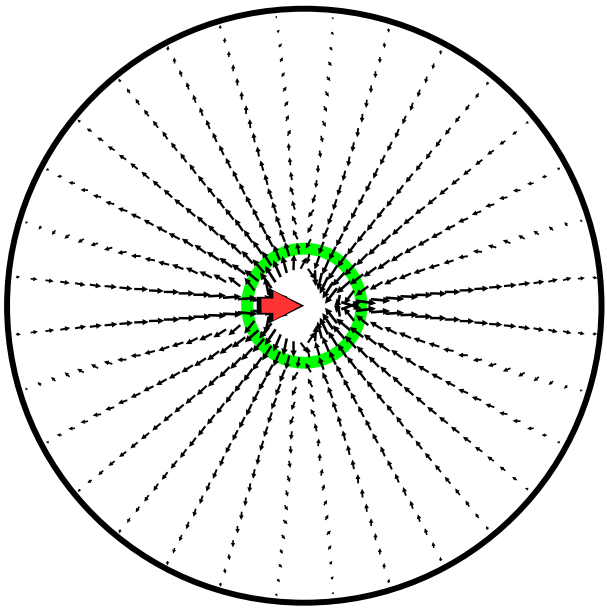}
\caption{{\em (Color online)} Mean displacement field after
averaging over  ${\mathcal O}(10^{5})$ individual fields induced
by imposing a displacement dipole at the origin. The orientation
of the dipole is given by the red arrow. As explained in the text,
the mean dipole moment vanishes after averaging and the resulting
field is quadrupole. A green circle at a radius $\lambda=0.191L$
from the origin is also shown. } \label{f:eg_quadrupole}
\end{figure}

\section{Results}
\label{s:results}

The mechanical response to a localized perturbation depends on the distance from the point of perturbation.
We divide the discussion
of these results into the following parts. First, we examine the response
to a point force using the decomposition of the displacement field outlined
in previous section. We contrast the decay of the non-continuum modes of
the displacement field with the behavior of the continuum modes of the
displacement field and discuss the finite-size effects in the simulations.
At large length scales, we find that the deformation field
approaches a quasi--continuum form, in which all but a small
number of ensemble--average modes decay rapidly toward zero. The
remaining modes of the deformation field are the same as those
predicted by continuum elasticity. In other words, the strain
field about a point force reflects the expected tensorial
character and rotational symmetries based on continuum elasticity
theory. We find that $\lambda$ again plays a central role in
controlling the cross-over from the near-field to the
quasi-continuum.

We then turn to the spatial structure of the elastic
energy density field around the point force paying particular
attention to the partitioning of that energy density between
stretching and bending modes of the filaments.
We observe that the ratio of these energy contributions
achieves the bulk value over much shorter distances from the
point force than does the structure of the displacement field
acquire its far-field, or bulk structure. 
We then extend our analysis to consider force dipoles in the medium.
Lastly, we extend our previous analysis of the bulk elasticity of the filament
network by examining  both the Young's modulus and the
Poisson ratio of the network.

\subsection{The response to point forces}
\label{ss:point-force}

\subsubsection{The Displacement Field}
\label{sss:displacement}

In this section we focus on the short length scale behavior of the
monopole response, for which the non--continuum modes are
non--zero. We wish to distinguish two distinct forms of this
non--continuum behavior: {\em (i)}~higher angular modes
$g^{(r)}_{m}$ and $g^{(f)}_{m}$ with $m>0$ are non--zero, and {\em
(ii)}~the $g^{(n)}_{m}$ modes do not vanish, {\em i.e.} the
response depends on the orientation of the filament to which the
force is applied. This latter observation gives a clear indication
of how the response can `see' the microscopic structure of the gel
on short length scales.

\begin{figure}[htpb]
\centering
\includegraphics[width=8.5cm]{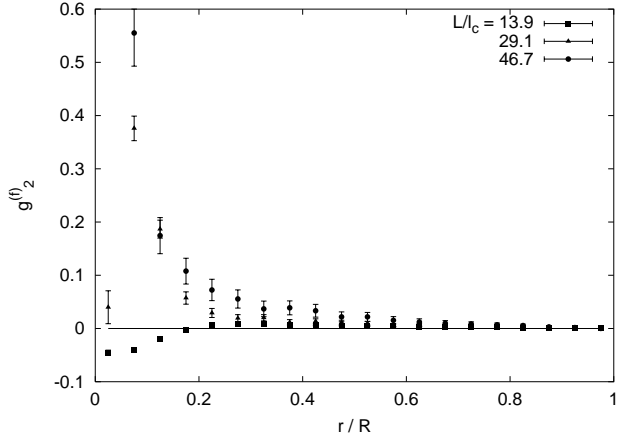}
\caption{$g^{(f)}_{2}$ (which is dimensionless) versus distance
from point of force application $r$ for systems with the same
$\lambda/L\approx0.191$ and crosslink densities $L/l_{\rm c}$ as
given in the key ($L$ is the filament length). The system radius
$R=L$ in all cases. } \label{f:gf2_5p2_R1L}
\end{figure}

An example demonstrating the appearance of non--continuum modes at
short lengths is given in Fig.~\ref{f:gf2_5p2_R1L}, which shows
the $g^{(f)}_{2}$ mode for systems with different crosslink
densities $L/l_{\rm c}$ but with the filament flexibility chosen
to give the same $\lambda\approx0.191L$ in each case. This plot
shows the decay of the $\cos (2 \theta)$ amplitude of the
component of the displacement field in the direction of the
applied force (at the origin). In all three systems one observes a
rapid decay of this angular harmonic that must vanish for a
continuum isotropic system. Moreover, the characteristic length
scale for this decay appears to be of order $\lambda$
($\simeq0.2R$), although we examine this point more quantitatively
below. For all network densities and values of $\lambda$ studied
it is clear that the magnitude of $g^{(f)}_{2}$ vanishes rapidly
with distance from the point force. No data are shown for larger
values of $L/l_{\rm c}$ since at high network densities the
numerical convergence of the strain field is so slow as to prevent
attaining meaningful statistics.

\begin{figure}[htpb]
\centering
\includegraphics[width=8.5cm]{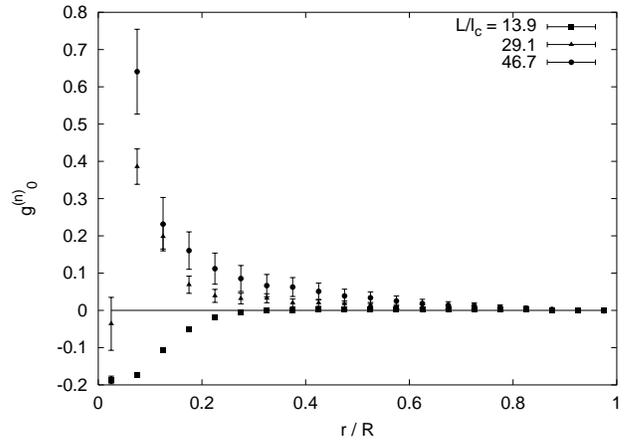}
\caption{$g^{(n)}_{0}$ versus $r/R$ for $\lambda/L\approx0.191$,
$R=L$ and the $L/l_{\rm c}$ given in the key. }
\label{f:gn0_5p2_R1L}
\end{figure}

This picture of non--continuum modes decaying to near zero at a
length comparable to $\lambda$ holds also for the $m>2$ modes of
$g^{(f)}_{m}$ and for all the $m>0$ modes of $g^{(r)}_{m}$. For
reasons of space we do not show these data here. Instead we
present data in Fig.~\ref{f:gn0_5p2_R1L} that demonstrate the
decay of $g^{(n)}_0$ with $r$ for three networks densities such
that in each case $\lambda \approx 0.191 L$.  This non--continuum
mode of the displacement field measures the circularly averaged
amplitude of the component of displacement in the direction
$\hat{n}$, {\em i.e.} along the axis of the rod to which the force
has been applied. Clearly, such behavior has no counterpart in an
isotropic continuum elastic model.  It is interesting to note that
this amplitude also appears to decay exponentially with a decay
length of order $\lambda$.

\begin{figure}[htpb]
\centering
\includegraphics[width=8.5cm]{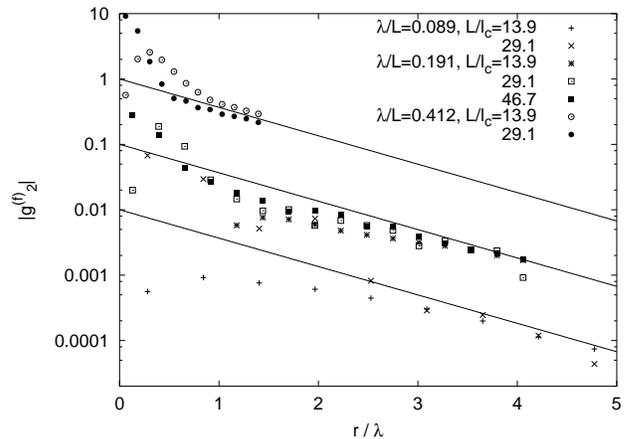}
\caption{Log linear plots of $g^{(f)}_{2}$ versus the scaled
distance from perturbation $r/\lambda$ for systems with various
values of $\lambda$ as given in the key.  The system radius $R=L$
in all cases. Data within a boundary layer of width $\lambda$
from $r=R$ is not shown (see text for details).}
\label{f:log-lin-gf2}
\end{figure}

In order to test quantitatively whether $\lambda$ indeed controls
the decay of those components of the displacement field that have
no counterparts in the continuum theory, we examine, as an
example, $g^{(f)}_2$ vs. $r$ more closely in
Fig.~\ref{f:log-lin-gf2}. Here we plot this for a range of values
of networks density {\em and} of $\lambda$. If, as suggested
above, the decay is exponential with characteristic length
$\lambda$, then plotting these data log-linear with radial
distances scaled by $\lambda$ should cause all these curves to
exhibit the same slope. We have shifted the data sets vertically
in order to facilitate visual comparison. (We note that, although
a known force is applied to the origin of our sample, since we
expect the unknown local constitutive relations in our
system to depend on density and other parameters, the amplitudes
cannot be directly compared. Here, we wish only to establish the
nature of the decay of the non--continuum modes.) We have also
introduced the solid lines corresponding to $~\exp(-r/\lambda)$
merely as guides to the eye. These are not fits, although fits to
these data for various $L/l_c$ and $\lambda$ demonstrate decay
lengths of $\lambda$ within 10\%.

Thus, we observe a near field regime in which the decay of the
$g^{(f)}_2$ appears to be generically more rapid than
$\exp(-r/\lambda)$ followed by the quasi-continuum regime where
the exponential decay with decay length $\lambda$ is observed.
This demonstrates that the nonaffinity length $\lambda$, indeed,
controls the approach toward the expected continuum behavior
(\emph{i.e.}, vanishing of the non--continuum modes). Similar
results (not presented here for reasons of space) are also found
for the other non--continuum modes. Specifically, the decay
lengths are also found to be $\lambda$ within 10\%.

Of course, the fixed zero-displacement boundary condition at $r=R$
requires all components of the displacement field to vanish there.
The essential distinction between the continuum and non-continuum
modes lies in the manner in which their amplitudes decay upon
approach to the rigid boundary.  The non-continuum modes examined
above decay exponentially with a decay length proportional to
$\lambda$. We demonstrate in the next section that the continuum
mode amplitudes, in contrast, do not appear to decay in the same
way. In particular, their approach to zero at the rigid boundary
is \emph{not} controlled by $\lambda$.

\subsubsection{Continuum modes and finite--size effects}
\label{sss:fintesize}

In this section we address two related points: (i) the fundamental
change in the structure of the displacement field as one moves
away from the immediate vicinity of the applied force, and (ii)
the role of finite size effects in our numerical simulation in
two-dimensions. We cannot address the former point without
confronting the latter one for the following reason. Due to the
presence of the rigid boundary, all modes of the displacement
field decay to zero at the boundary.

We show in this section, however, that the decay of the
non-continuum modes is controlled by internal, mesoscopic length
scales in the network, while the decay of the continuum modes is
determined by the macroscopic geometry of the system including the
presence of the rigid boundary. Since the distinction between the
continuum and non-continuum modes depends essentially on the
presence of the boundary and specifically on the separation of the
system size $R$ from the internal length scales controlling the
decay of the non-continuum modes ($\propto \lambda$), we consider
this distinction along with a more general discussion of finite
size effects in our simulation.


\begin{figure}[htpb]
\centering
\includegraphics[width=8.5cm]{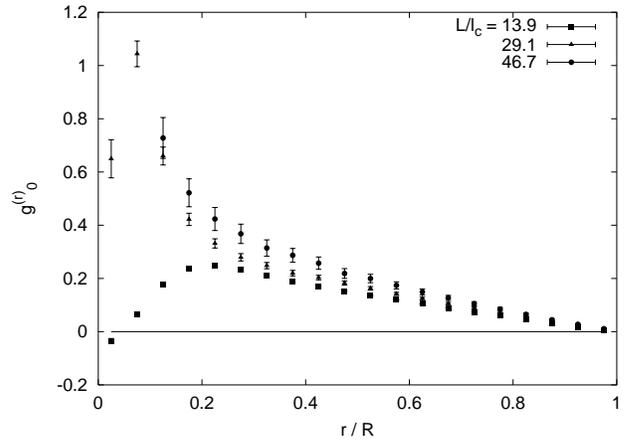}
\caption{$g^{(r)}_{0}$ versus $r/R$ for $\lambda/L\approx0.191$,
$R=L$ and differing $L/l_{\rm c}$ as shown in the legend. }
\label{f:gr0-R=L}
\end{figure}


To have a well-defined elastic response in two-dimensions (meaning
that the displacement field vanish at large distances from the
applied point force) one needs to impose a rigid boundary. In
order to eliminate the differential influence of the boundary on
the various angular modes of the displacement field we chose this
boundary to be a circle of radius $R$ centered on the point of
force application. This rigid boundary forces all components of
the displacement field to vanish exactly at $r=R$. Such a rigid
boundary can be expected to introduce a boundary layer near the
edges of our system, which we observe and discuss below.

Figure \ref{f:gr0-R=L} shows the decay of the $g_0^{(r)}$ mode amplitude
for systems of size $R=L$ and $\lambda\approx0.191L$.
We note that the decay of this \emph{continuum} mode is
qualitatively distinct from the decay of the non-continuum modes
shown in Figs.~\ref{f:gf2_5p2_R1L}, \ref{f:gn0_5p2_R1L}, and
\ref{f:log-lin-gf2}. We do not observe an exponential decay of
this continuum mode amplitude. Nevertheless, observing an
exponential decay having a long decay length, or, more reasonably,
the product of an algebraic function and such a weak exponential
decay would be difficult to resolve in this plot.

\begin{figure}[htpb]
\centering
\includegraphics[width=8.5cm]{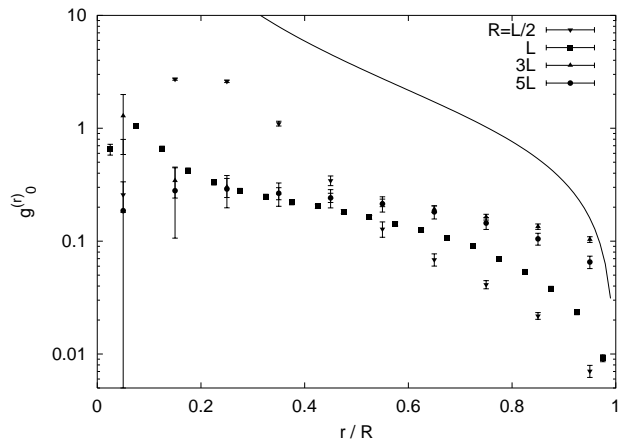}
\caption{$g^{(r)}_{0}$ versus $r/R$ for $\lambda/L\approx0.191$,
$L/l_{\rm c}\approx29.1$ and the system radii $R$ given in the key.
The continuum response for the bulk elastic moduli corresponding to
these parameter values (as given by (\ref{e:ctm_gr0})) is shown as a solid line.
}
\label{f:gr0_syssize}
\end{figure}

To study these issues further, we plot in Fig.\
\ref{f:gr0_syssize} $g^{(r)}_0$ for various system sizes from
$R=L/2$ to $5L$. We observe both an apparent convergence for the
larger systems to a common curve when distances are measured
relative to the size of the system, as well as a systematic
down-turn near the boundary. The first of these observations
suggests that the sample geometry controls the dependence of this
continuum mode, as opposed to intrinsic lengths like $\lambda$.
The down-turn in these data within a distance of order $\lambda$
(here, $\simeq 0.2L$) of the boundary is to be expected from the
observations above concerning the role of the length scale
$\lambda$. One can view this crossover length as the distance
along a filament over which a force applied to the filament
dissipates/expands into a (quasi-)continuum stress/displacement
field. Thus, the effects of the boundaries  on the individual
filaments contacting the boundary are expected to propagate at
least this distance into the system. Because of this, we have also
removed all data within a distance $\lambda$ of the boundaries in
Fig.\ \ref{f:log-lin-gf2} above, which exhibit a similar down-turn
near the boundary.

\begin{figure}[htpb]
\centering
\includegraphics[width=8.5cm]{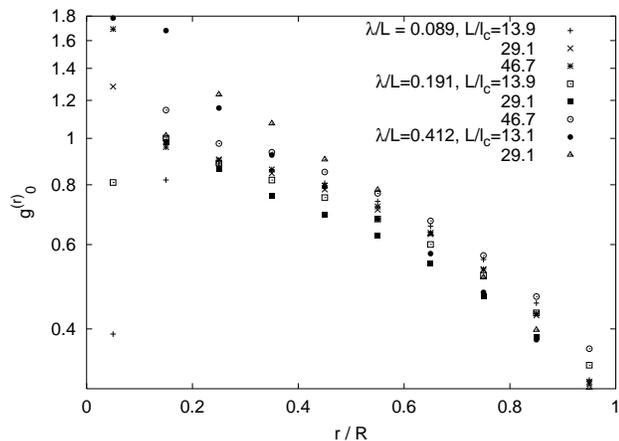}
\caption{$g^{(r)}_{0}$ versus $r/R$ for $R=3L$ on log--linear axes, with
$\lambda/L$ and $L/l_{\rm c}$ as given in the key. Each curve has been shifted
vertically by a $\lambda$--dependent scale factor to attempt data collapse.
Since $\lambda$ varies by over a factor of 4, the rough collapse is enough to
rule out any significant $\lambda$--dependence on the decay.
Error bars are not shown for clarity.}
\label{f:lambda_collapse}
\end{figure}

In Fig.~\ref{f:lambda_collapse} we examine the radial dependence
of $g^{(r)}_0$ for three different values of $\lambda/L$ for a
large system where $R = 3 L$.  In this case, we see a coincidence
of the data for different $\lambda$, indicating that the
long-range behavior of $g^{(r)}_0$ is not controlled by $\lambda$.

Although we were unable to obtain sufficient statistics on the
non--continuum (and sub--dominant) modes for larger systems, the
combination of the qualitatively different behavior from the
continuum modes, together with the consistent dependence on the
crossover length $\lambda$ lead us to conclude that the principal
results of the prior section are not strongly influenced by
finite-size effects. Thus, $\lambda$ controls the disappearance of
the non-continuum modes of the system.

We also see a large discrepancy between the expected continuum
solution and the observed displacement of the simulated filament
network as shown in Fig.\ \ref{f:gr0_syssize}. This difference
cannot be simply attributed to finite-size effects; as can be seen
in Fig.~\ref{f:gr0_syssize}, there is no observable convergence of
the larger system data for the continuum modes to the predictions
of continuum elasticity for a material having the appropriate
Lam\'{e} coefficients. Nevertheless, we clearly observe the more
rapid decay of the non-continuum modes on a length-scale
controlled by $\lambda$. One is left with the following puzzle:
The tensorial structure and the rotational symmetry of each
component of the displacement field approaches the form required
by the continuum theory, but the continuum model appears to never
quantitatively agree with the numerical data.

We may speculate about the underlying cause of this discrepancy.
It is clear that that the deformation of the material in the
immediate vicinity of the point force (this ``near zone'' extends
out to a distance $\propto \lambda$ from the point force) and in a
boundary region at the rigid wall is not well described by any
continuum theory. We suggest that under the application of the
point force at the origin, the network effectively partitions
itself into three different elastic materials. In the near field
region $r < \lambda$ around the point force the deformation
response to the point force is quite complex. Similar complexities
appear to exist near the rigid wall, where $R- r < \lambda$ .
These regions apply a complex set of tractions on the circles $r
\approx \lambda$ and $r \approx R-r$  bounding the intermediate
region that deforms in a manner consistent with some continuum
elastic material. Because these tractions are not themselves
determined by a simple, continuum model, the resultant deformation
of the intermediate region is also not simply derivable from an
analysis of the elastic Green's function for a continuum.

One might imagine that for very large systems having a
consequently larger intermediate region, the complexities of the
tractions in the transition zones become less significant.
Because of the rigid boundary at $r=R$ and the low-dimensionality
of the system, this may not be the case. The continuum Green's
function contains logarithmic terms and, due to the rigid
boundary, growing polynomial terms as well. Thus, we do not expect
convergence to the continuum Green's function even for
significantly larger systems. One may ask whether three
dimensional systems will show similarly poor convergence to the
continuum solutions. Further research here is needed.

We suggest that we do indeed observe the approach of the structure
of the elastic Green's function to that predicted by continuum
theory.
We term the region surrounding the point force where the
deformation field has the expected form the quasi-continuum.
Based on our numerical data, we do not expect to find a region in
which the deformation field agrees quantitatively with the
predictions of continuum elasticity using the Lam\'{e}
coefficients appropriate to the medium as determined by uniform
stress measurements. This suggests that semiflexible networks
admit a highly complex point force response that cannot be fully
captured by continuum elasticity even in the far-field. The full
implications of this complexity have not been explored.

\subsubsection{Energy Density}
\label{sss:energy-density}

Another instructive measure of the monopole response is the
density of elastic energy $\rho_{\rm E}$ at a given distance from
the point perturbation, which measures gradients of the
displacement field and therefore complements the analysis of ${\bf
u}({\bf x})$ given above. In addition, since the simulation
contains explicit terms for the transverse and longitudinal
filament deformation modes, it is straightforward to
measure the
partitioning of the energy between these modes. Based on previous
work \cite{head:03a,wilhelm:03,head:03c}, under homogeneous shear
strain the partitioning of elastic energy between the bending and
stretching modes of the filaments is determined entirely by the
affine-to-nonaffine crossover. The ratio of $L/\lambda$ was found
to control this energy partitioning at a macroscopic or average
level. It remains to be seen how this partitioning of the elastic
energy occurs in vicinity of a point force.

The freedom to choose the angle between the point force and the
direction of the filament to which that force is applied allows
one to determine locally the partitioning of the elastic energy
between bending and stretching modes of the filaments. Forces
directed along the filament axis generate primarily stretching
deformations in the immediate vicinity of the origin (where the
force is applied). Forces directed perpendicular to the filament
axis, however, locally create a large bending deformation. As seen
in the previous homogenously imposed strain deformation
calculations, for any given value of $L/\lambda$ the network
responds very differently to the bending or stretching
deformations. Thus it is not surprising that the decay of the
energy density from the point of force application to the boundary
is strongly dependent on whether the force is applied parallel or
perpendicular to one of the filaments at the crosslink.

\begin{figure}[htpb]
\centering
\includegraphics[width=8.5cm]{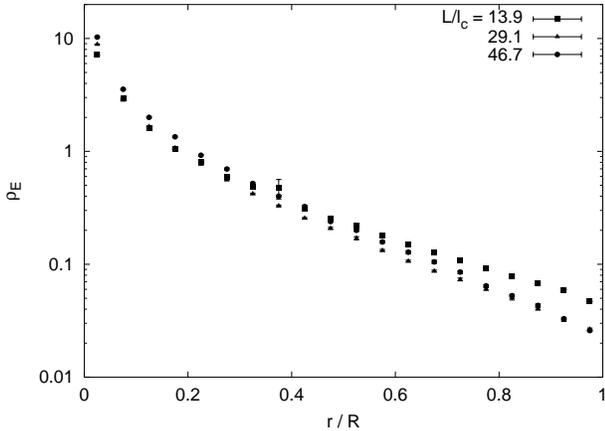}
\caption{Density of elastic energy $\rho_{\rm E}$ (in arbitrary
units) versus distance $r$ from an externally applied force that is directed along the filament to which it is applied. $L/l_{\rm c}$ is varied as given in the key, and $R=L$ and
$\lambda/L\approx0.191$ in all cases. }
\label{f:totED_fixedlam_par}
\end{figure}

Fig.~\ref{f:totED_fixedlam_par} shows $\rho_{\rm E}$ for parallel
forces, where $\lambda$ is fixed but the filament density
$L/l_{\rm c}$ varies. There is an approximate data collapse onto a
single curve as shown. The most notable discrepancy occurs near
the boundary, where the low--density data remains higher than
those for larger $L/l_{\rm c}$ values. This is most likely due to the boundary
layer already discussed above for the displacement modes.

\begin{figure}[htpb]
\centering
\includegraphics[width=8.5cm]{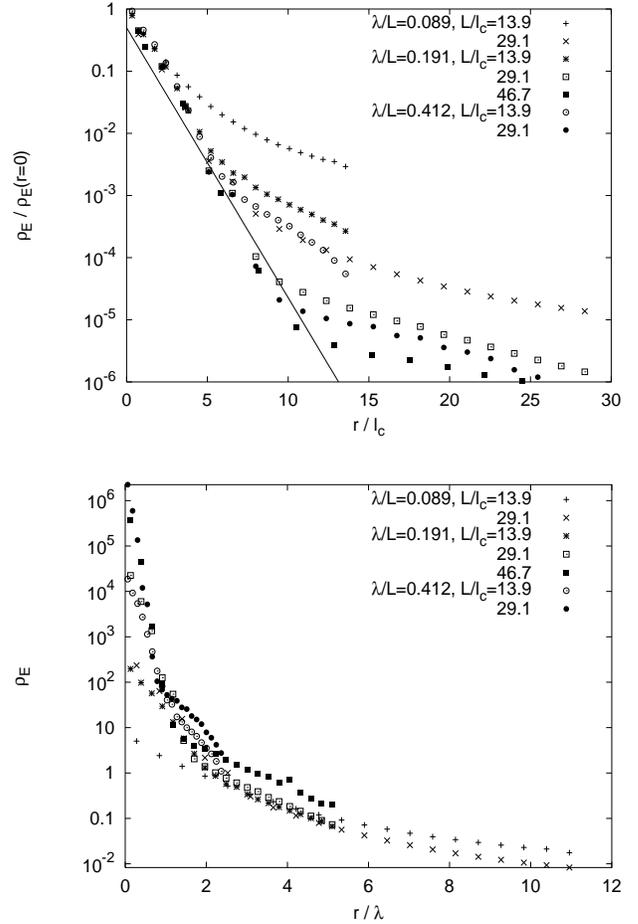}
\caption{Scaled elastic energy density versus distance from the
point force for a variety of networks -- see legend. In each case
the force is directed perpendicularly to the filament where it is
applied. In the upper panel distances are scaled by $l_{\rm c}$ and the
data has been collapsed in the small $r/l_{\rm c}$ regime. The solid
line is proportional to $\exp(-r/l_{\rm c})$,
showing $l_{\rm c}$ controls the initial decay of elastic energy.
In the lower panel the same energy densities are plotted against
distance scaled by $\lambda$. This shows the energy density decays
with that characteristic length scale at longer ranges. Error bars
are not shown for reasons of clarity. }
\label{f:totED_fixedlam_perp}
\end{figure}

In contrast, $\rho_{\rm E}$ for forces perpendicular to a filament
exhibits much richer behavior, as shown in
Fig.~\ref{f:totED_fixedlam_perp}, which shows the same data
plotted against $r/l_{\rm c}$ and $r/\lambda$. For large
distances, the data for different $L/l_{\rm c}$ appear to differ
by only a scale factor. Since there is one arbitrary
multiplicative factor for each curve (as only the magnitude of the
applied force is fixed, not its displacement and hence work done
on the network), these portions of the curves can be made to
collapse after scaling, as in the parallel force case, suggesting
that beyond some near-field regime, the decay of elastic energy
density is once again exponential and governed by $\lambda$ -- see
Fig.~\ref{f:totED_fixedlam_perp} lower panel. However, in this
near-field regime for perpendicular forces there is clearly no
possibility of such a collapse. Instead, it appears that the more
rapid decay of elastic energy density in this near-field regime is
governed by the length $l_{\rm c}$, as can be seen in the upper
panel of Fig.~\ref{f:totED_fixedlam_perp} where the (rescaled)
energy density data are plotted vs. $r/l_{\rm c}$ and shown to
decay as $\sim\exp(-r/l_{\rm c})$ for distances
$r\stackrel{<}{\scriptstyle\sim}5l_{\rm c}$ (ignoring finite size
effects).

Based on the two different regimes of data collapse shown in
Fig.~\ref{f:totED_fixedlam_perp}, we conclude that there is a
near-field region in which the decay of elastic energy is governed
by a microscopic length scale--the mean distance between
cross-links--and a longer-range regime in which the spatial decay
of elastic energy density is controlled by the mesoscopic length
$\lambda$. To better study this cross-over, we examine the
partitioning of elastic energy between bending and stretching
modes of the filaments. Recall that in the affine limit, the
elastic energy is stored primarily in the stretching and
compression of the filaments. In the nonaffine regime, on the
other hand, the elastic energy is stored almost entirely in
bending modes of the filaments.

\begin{figure}[htpb]
\centering
\includegraphics[width=8.5cm]{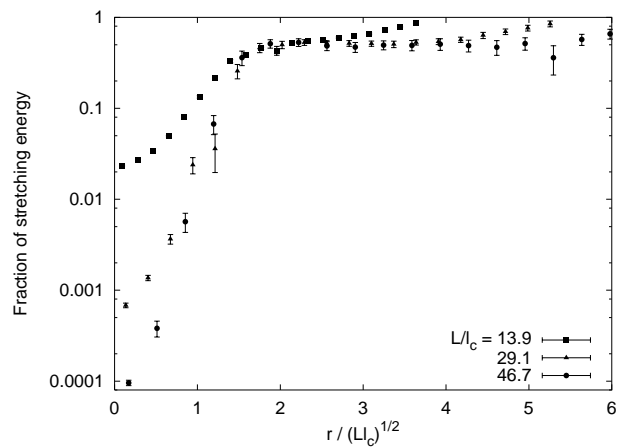}
\caption{Proportion of elastic energy due to filament stretching,
${\cal H}^{\parallel}/({\cal H}^{\parallel}+{\cal H}^{\perp})$,
for three systems having differing values of $l_c$ (see key) but
the same value of $\lambda$. } \label{f:fracED_fixedlam_perp}
\end{figure}

\begin{figure}[htpb]
\centering
\includegraphics[width=8.5cm]{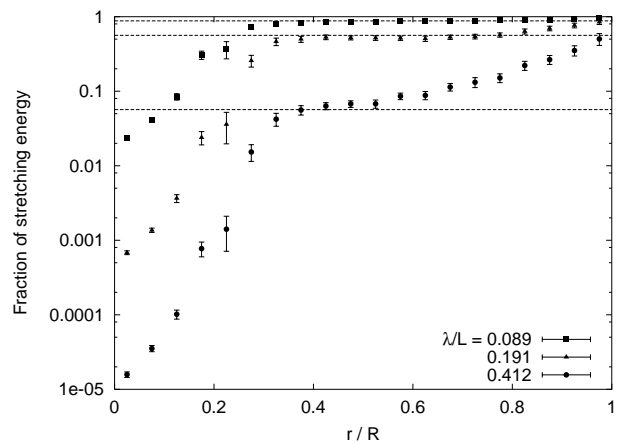}
\caption{The fraction of elastic energy due to filament stretching
in networks with the same $L/l_{\rm c}\approx29.1$ (in fact, the
exact same geometries), $R=L$ but different~$\lambda$. The
different values of the stretching energy fraction in the
intermediate regime reflect the differing values of the ratio
$L/\lambda$ for the networks. For comparison, the macroscopic energy fraction
under bulk shear is shown as horizontal dashed lines, in the same order as the
data points ({\em i.e.} $\lambda/L\approx0.089$, 0.191 and 0.412 from top to bottom).
}
\label{f:fracED_fixedlc_perp}
\end{figure}

Plotted in Fig.~\ref{f:fracED_fixedlam_perp} is the proportion of
elastic energy due to longitudinal filament deformation for the
perpendicular force case. Three data sets are displayed having
different values of $l_c$ and $\lambda$ -- see figure caption. The
distance from the point of force application has been scaled by
the geometric mean of $l_c$ and $L$ leading to the observed
coincidence of the crossover between regimes in the data. Close to
the point of the application of the force, the displacement
response is clearly dominated by bending modes of the filaments.
This is to be expected since the perpendicularly directed force
directly injects bending energy into the system at the origin --
the point of force application. The energy partitioning is
inconsistent with that which is expected based on previous
homogeneous shear measurements. From that work, one expects the
network to determine the energy partitioning based only on the
ratio: $L/\lambda$.  We observe the fraction of stretching energy
to rapidly increase towards the expected value based on
$L/\lambda$ \cite{head:03a,head:03c} and we may characterize each
curve has having a `knee' separating the region of varying energy
ratio from a nearly constant intermediate regime where the
partitioning of elastic energies corresponds well with the
previously identified fraction of stretch/compression energy under
macroscopic strain. This correspondance is demonstrated in
Fig.~\ref{f:fracED_fixedlc_perp}. The horizontal lines show the
fraction of stretching energy in periodic systems subjected to macroscopic
shear, and clearly coincide with the plateau reached beyond the knee.
At the largest distances one notices the vanishing
of bending energy as the fraction of stretching energy approaches
unity. This last effect is due to the fact that each filament has
a freely rotating bond at the outer, rigid wall. As this wall
cannot support torques, the bending energy vanishes in a boundary
layer whose width is determined by the mean distance between
cross-links. We return to this point below.

The observed coincidence of the knees of all three curves under
the rescaling of distances by $\sqrt{l_c L}$ suggests that this
length sets the scale over which injected bending energy is
redistributed into the combination of bending and stretching
appropriate for long-length scale deformations.

Based on these observations we note that in all networks the
strain field acquires the structure of the point force response
based on continuum elasticity over a (typically mesoscopic) length
scale of $\lambda$. When the networks is subjected to large, local
bending deformations, however, it readjusts the partitioning of
bending to stretching energy over a generally much shorter length
scale, $l_c$.  Thus the system is able to repartition the local
elastic energy storage to the value appropriate for its
$L/\lambda$ ratio over smaller length scales than does the system
recover the expected long length scale structure of its continuum
elastic response.

\begin{figure}[htpb]
\centering
\includegraphics[width=8.5cm]{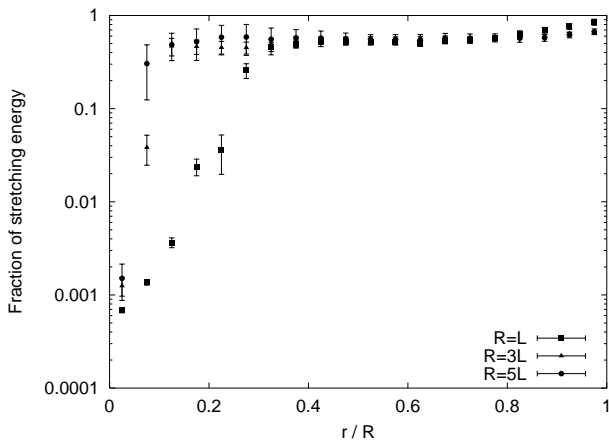}
\caption{The fraction of elastic energy stored in the ${\cal
H}^{\parallel}$ term of the Hamiltonian for perpendicular forces,
as a function of distance from the origin~$r$ for system sizes
$R=L$, $3L$ and $5L$. In all cases $L/l_{\rm c}\approx29.1$ and
$\lambda\approx0.191L$. } \label{f:fracED_syssize_perp}
\end{figure}

We have already shown that in the intermediate field regime the
fraction of stretching elastic energy in the system approaches its far-field value as determined by the ratio: $L/\lambda$.
Nevertheless, we in fact find that at the edges of our sample the
network energy becomes stored solely in stretching modes
regardless of the value of $L/\lambda$. As mentioned above we
attribute this final redistribution of the elastic energy density
between bending and stretching modes to a boundary effect imposed
by the freely-rotating nature of coupling of the network filaments
to the rigid boundary. To further test that this final
redistribution is indeed a boundary effect, we consider a few
larger system sizes as shown in Fig.~\ref{f:fracED_syssize_perp}.
In that figure the force remains perpendicular to the direction of
the filament to which it is applied while the system size is
varied from $R = L$ to $R= 5 L$ for a network of fixed $\lambda$.
When the data are plotted against the radial distance from the
point force scaled by system size, we find an excellent collapse
in the intermediate regime and in the putative boundary layer. For the smallest system size
considered, $R=L$, we note poorer data
collapse in the near field region suggesting that one must study
systems that are are least larger than a single filament length to access the bulk behavior of the network
with quantitative accuracy. Clearly, all three curves taken together are consistent with the
notion of an elastic boundary layer that is produced by the freely-rotating boundary at the
wall and that extends distance approximately equal to $\lambda$ into the sample.

\subsection{Network response to force dipoles}
\label{ss:dipole}

\begin{figure}[htpb]
\centering
\includegraphics[width=8.5cm]{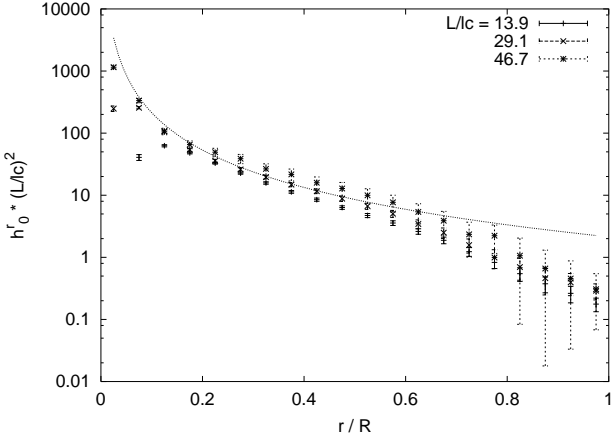}
\caption{$h^{r}_{0}$ for dipole forcing for fixed
$\lambda/L\approx0.191$ compared to the continuum solution (smooth
line), which has one free fitting parameter, namely the overall
magnitude of the dipole forcing. Both data and curve are negative,
so the magnitudes have been plotted to allow use of a logarithmic
axis. Each data curve has been scaled by $(L/l_{\rm c})^{2}$ to
ensure the same mean dipole magnitude (see text). The system
radius was $R=L$. } \label{f:dip1}
\end{figure}

We now consider the mechanical response of the network to
localized force dipoles at the origin. Understanding this response
function is central to elucidating the effect of molecular motor
activity in the cytoskeleton. Many of the general features
observed in the response of the system to point force monopoles
are also in evidence. For example in Fig.~\ref{f:dip1} we see the
rough collapse in the far field of the mode amplitude $h^r_0(r)$
to the continuum solution for three different networks having the
same value of $\lambda$. Note that the amplitude has been scaled
by $l_{\rm c}^2$ since the magnitude of the imposed force dipole
for each realization of the network will depend on the distance to
the next constraint, {\em i.e.} cross-link, which is $l_{\rm c}$.
Since the displacement field when averaged over network
realizations is quadrupolar, being the difference in two dipole,
that length must enter squared.  The analogous displacement field
amplitude, $h^r_0(r)$ from the continuum solution (solid line in
Fig.~\ref{f:dip1}) was calculated using unit forces so this
amplitude is known only up to an overall scale factor; that scale
factor was adjusted to best fit the data. A similar comparison can be made for the
other continuum modes of the displacement field (see the Appendix).
Fig.~\ref{f:dip2}, for example, shows $h^{f}_{0}(r)$
with the same arbitrary prefactor. The agreement to the continuum
theory scaled as discussed above is quite poor. As is also seen
in the point-force response of the strain field,
higher angular modes, which vanish in the
continuum theory such as $h^{f}_{4}$ are significantly non--zero in the data.
The amplitude of the non-continuum modes in the dipolar response is
even more dramatic than in the point-force response of the
network examined earlier.

The simulation data for the response of the network to force
dipoles can be broadly summarized as follows: The amplitude
of the continuum  modes of the dipolar displacement field from the
naive continuum theory do not agree with simulation data.
In light of the discussion regarding such disagreements between
the  monopole data and the continuum calculation, this is
perhaps not surprising. More interestingly, the observed
displacement field has significant amplitudes of higher order
angular modes. In short the universality of the response of
semiflexible networks to localized point forces does not
appear to extend to their response to localized force dipoles.

We speculate that the principal difference between these
two cases stems from the fact that the network's response to the force
dipole probes the more detailed microscopic structure of the
network in the immediate vicinity of the point of the force dipole
application. The amplitude of the lowest order force multipole
communicated from the near field region to the far field where we
expect a continuum based theory to apply is not constrained by
elementary force balance. Neither are the amplitudes of any higher
order force multipoles generated within the near field region.
Thus, upon reaching the inner edge of the far field region the
force dipole imposed at the origin has generated a highly complex
set of tractions on the rest of the material whose structure
depends on local details of the connectivity of the network near
the origin. In contrast for the case of the force monopole, the
dominant term in those tractions is the fixed total monopolar
force acting on the intermediate region. The higher order force
multipoles created in the inner region decay rapidly with distance
from the origin leaving one with highly reproducible results for
the response of the network to applied forces. The amplitude of
the force dipole communicated from the near field region to the
intermediate region, on the other hand, is not similarly
constrained. It appears that one generically generates large
amplitude higher order force multipoles in addition to whatever
force dipole is communicated to the intermediate region making
convergence to a simple dipolar form slow and difficult to observe
in our finite samples.

\begin{figure}[htpb]
\centering
\includegraphics[width=8.5cm]{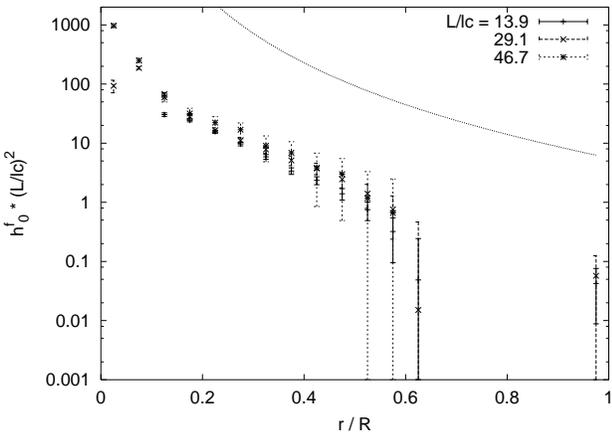}
\caption{$h^{f}_{0}$ for same systems as in Fig.~\ref{f:dip1}. The
only fitting parameter for the continuum solution is the same as
used previously for $h^{r}_{0}$, so there are no remaining free
fitting parameters in this plot. Much of the data for $r/R>0.6$ is
negative and hence not visible on these axes. } \label{f:dip2}
\end{figure}

\subsection{Bulk moduli}
\label{ss:bulk}

\begin{figure}[htpb]
\centering
\includegraphics[width=8.5cm]{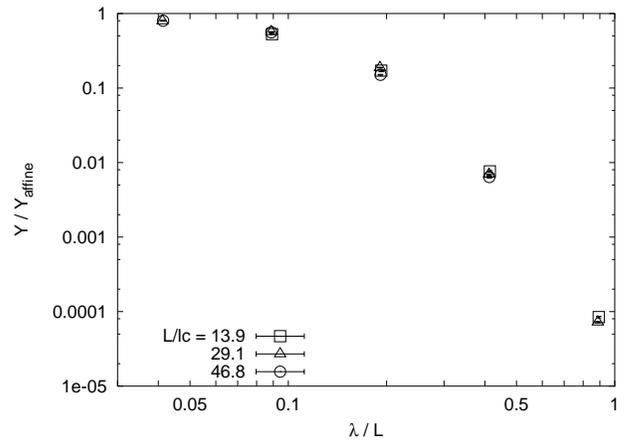}
\caption{Young's moduli scaled by the affine prediction versus
$\lambda/L$ on log--log axes. The affine prediction, which depends
only on $L/l_{\rm c}$, can be found in~\cite{head:03c}. The
symbols are larger than the error bars. } \label{f:Y}
\end{figure}

Lastly, we present new results on homogeneous deformations of
semiflexible networks. It has already been shown that the
macroscopic elastic moduli of this class of model networks depend
in a crucial way on the ratio of $\lambda$ to the filament
length~$L$~\cite{head:03a,head:03c}. For $\lambda/L\ll1$, the
deformation is approximately affine, whereas non--affine
deformation modes dominate when
$\lambda/L\stackrel{>}{\scriptstyle\sim}1$. Previously this was
demonstrated only for the shear modulus; here we can now confirm
that the Young's modulus $Y$ behaves in an identical manner.
Fig.~\ref{f:Y} shows $Y$ measured from uniaxial extension of a
rectangular cell, scaled by the prediction for an affine strain,
plotted against $\lambda/L$ for the range of $L/l_{\rm c}$
considered in this paper. There is a clear data collapse, as for
the shear modulus. Furthermore the deviation from the affine
prediction is small for $\lambda/L\ll1$, but becomes increasingly
pronounced as $\lambda/L$ increases. This confirms that
$\lambda/L$ controls the macroscopic elastic response of these
systems.

\begin{figure}[htpb]
\centering
\includegraphics[width=8.5cm]{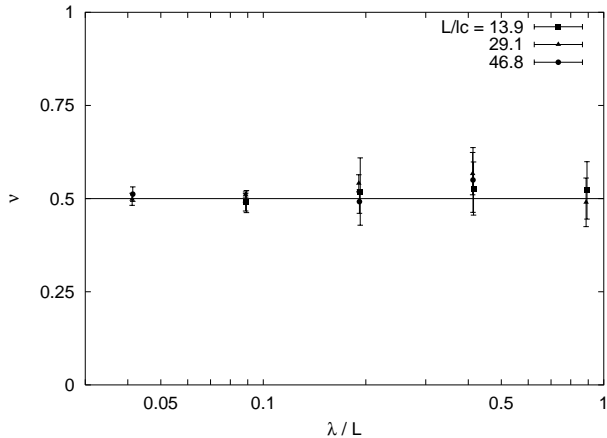}
\caption{The Poisson ratio $\nu$ for the same systems as in
Fig.~\ref{f:Y}. The solid line corresponds to $\nu=\frac{1}{2}$. }
\label{f:nu}
\end{figure}

Fig.~\ref{f:nu} gives the Poisson ratio $\nu$ for the same systems
as in Fig.~\ref{f:Y}. It is striking to observe that, within error
bars, $\nu$ is consistent with the value $\nu=\frac{1}{2}$, which
is the value expected for an affine deformation~\cite{head:03c}.
However, it is apparent from this figure that the measured values
are consistent with $\nu=\frac{1}{2}$ for {\em all} data points,
even those well into the non--affine regime. The mechanism behind
this striking robustness currently evades us (is has nothing to do
with incompressibility, which fixes $\nu=1$ in two dimensions).
Note that $\nu$ at the rigidity percolation $L/l_{\rm
c}\approx5.933$ (at which the elastic moduli vanish) is
$\approx\frac{1}{3}$~\cite{head:03b}, which is clearly
inconsistent with the data in Fig.~\ref{f:nu} and confirms our
earlier claims that the non--affine regime is distinct from the
scaling regime of the transition.

\section{Discussion}
\label{s:discussion}

In this paper we have presented the results of numerical studies
on the response of semiflexible networks to both point forces
and to homogeneously imposed strain. The data presented on the
Young's modulus taken in combination with previous work on the
static shear modulus shows that the mechanical response of the
system can be understood in terms of Lam\'{e} coefficients that
depend on the ratio of the filament length to the nonaffinity length: $L/\lambda$.
The Poisson ratio of the material, however, appears to be remarkably insensitive
to this ratio. We can offer no explanation for this insensitivity at this time.
The data presented on the point force response form a necessary compliment to
previous work on the development of a long length scale elastic theory of such materials.

Based on these data, it appears that the storage of
elastic energy and the structure of the strain field
is rather complex in the immediate vicinity of the
applied point force. We characterize these quantities by
considering three qualitatively different regimes as a
function of radial distance from the applied point force.
Immediately surrounding the point force in the near-field
regime we find that the partitioning of elastic energy
into bending and stretching is determined primarily
by the angle between the applied force and the direction
of the filament to which that force is applied. The
disorder-averaged strain field is rather complex
having higher angular harmonics that predicted
by continuum elasticity theory. In the intermediate
field regime farther from the point force the
partitioning of elastic energy is determined
solely by the ratio $L/\lambda$ as found in
the homogeneous shear measurements. The higher
angular harmonics present in the strain field appear
to decay exponentially with a decay length proportional
to $\lambda$ with a constant of proportionality near unity.
These results taken in combination with our previous work
suggests that one may think of $\lambda$ as setting the
minimum length scale for the applicability of continuum
modeling quite generally.

In the quasi-continuum regime ($r > \lambda$),
we find a strain field consistent with structure
of that predicted by continuum model. The strain
itself, however, cannot be simply computed from a
knowledge of the effective Lam\'{e} coefficients.
We believe the source of this discrepancy is the
fact that the intermediate region, being poorly
described by the continuum theory, applies a much
more complex set of tractions to the material in
the far-field where the continuum theory must apply.
If these boundary conditions were known, we suspect
that one could in fact calculate the resulting
displacement field using the continuum theory. This
belief is  supported by the fact that in the far-field
regime the displacement field for different networks
having the same $\lambda$ collapse onto the same
curve as one would expect for any effective
continuum theory in which the Lam\'{e} coefficients depend on $\lambda$.

To summarize these results, we suggest that there
is an emerging
description of the mechanics semiflexible networks.
There mechanical behavior over length scales longer
than $\lambda$ appears to be described by a modified
 elastic theory in which the effective Lam\'{e}
 constants depend on the ratio $L/\lambda$.  For
 point forces and presumably any force applied
 over regions having a characteristic length
 scale that is small compared to $\lambda$,
 the local response of the network is quite complex
 and the material appears to be anomalously compliant;
 at longer length scales, however, the structure
 of the deformation field appears to be consistent
 with the predictions of continuum elasticity.
 We believe that it should be possible to construct
 an elastic continuum theory of these networks that
 is applicable in both the intermediate and far-field
 and that is based on a modified gradient expansion of
 the strain field incorporating explicitly the mesoscopic
 length $\lambda$.  The behavior of the strain field on
 scales much smaller than $\lambda$ appears to depend on
 other, more microscopic length scales.

Understanding the response of semiflexible networks to localized
forces is a necessary for both microrheological investigations
of semiflexible networks and for understanding the effect of
 molecular motors in the cytoskeleton. Clearly, further
 numerical investigations are required as well as a
 theoretical examination of the development of elastic
 continuum models applicable to the intermediate and
 far-field regimes.

Moreover, additional investigations are required to examine the
analogous questions in three dimensional semiflexible networks.
While we expect the basic physics outlined above, including the
existence of a mesoscopic length $\lambda$, to persist in three
dimensions, the expected $1/r$ decay of the elastic Green's function
should alter the results. Moreover the more rapid decay of the
displacement field and energy density in the continuum three
dimensional system may further simplify the structure of the
analogous Green's functions for the semiflexible network.

\section*{Acknowledgements}

AJL and FCM acknowledge the hospitality of the Isaac Newton
Institute for Mathematical Sciences where part of this work was
done. DAH was jointly funded by the European Community
Marie Curie and the Japanese Society for the Promotion
of Science programs.  AJL was supported in part by NSF-DMR0354113.

\appendix

\section*{Appendix}
\label{s:app_ctm_calcs}

Here we derive the displacement field ${\bf u}({\bf x})$ predicted
by continuum elasticity for the cases of both monopole and dipole
forcing. For the monopole case, a point force ${\bf f}\delta({\bf
x})$ is applied to the origin of an elastic sheet with Lam\'e
coefficients $\shear$ and $\lambda$, or equivalently $\shear$ and
the Poisson ratio~$\nu=\lambda/(2\shear+\lambda)$. For an
isotropic elastic body, the stress obeys
$\sigma_{ij}=\shear(\partial_{i}u_{j}+\partial_{j}u_{i})+\lambda\delta_{ij}
\mbox{\boldmath$\nabla$}\cdot{\bf u}$ and the resulting equation
for force balance is~\cite{landau:86}
\begin{equation}
(\shear+\lambda)\partial_{i}(\mbox{\boldmath$\nabla$}\cdot{\bf u})
+\shear\mbox{\boldmath$\nabla$}^{2}u_{i} = f_{i}\delta({\bf x})\,.
\end{equation}
Solving this in polar coordinates $(r,\phi)$ with the boundary
condition ${\bf u}\equiv{\bf 0}$ at a radius $R$ from the origin
eventually leads to
\begin{eqnarray}
u^{\rm mono}_{i}&=&\frac{f}{8\pi\shear} \left\{
4\hat{f}_{i}\ln(r/R)
-4c_{1}\hat{r}_{i}\hat{r}_{j}\hat{f}_{j}\right.
\nonumber\\
&&\mbox{}+c_{2}\left(r/R\right)^{-2}
\left[2\hat{r}_{i}\hat{r}_{j}\hat{f}_{j}-\hat{f}_{i}\right]
\nonumber\\
&&\left.\mbox{}-\left(r/R\right)^{2}
\left[2\hat{r}_{i}\hat{r}_{j}\hat{f}_{j}-c_{2}\hat{f}_{i}\right]
\right\}\,,
\end{eqnarray}
where $c_{1}=(1+\nu)/(3-\nu)$ and $c_{2}=(5+\nu)/(3-\nu)$. Then
the only two non--zero modes according to the definition of the
$g^{(\cdot)}$ given in~(\ref{e:g_modes}) are
\begin{eqnarray}
g^{(r)}_{0} &=& \frac{\shear_{\rm aff}}{8\pi\shear} \Big\{ -4c_{1}
+2c_{2}(r/R)^{-2} -2(r/R)^{2} \Big\}, \label{e:ctm_gr0}
\\
g^{(f)}_{0} &=& \frac{\shear_{\rm aff}}{8\pi\shear} \Big\{
4\ln(r/R) -c_{2}(r/R)^{-2} +c_{2}(r/R)^{2} \Big\}.
\nonumber\\
\end{eqnarray}

A naive calculation of the corresponding dipole solution would
simply superpose the above monopole solution for two point forces,
${\bf f}\delta({\bf x})$ and $-{\bf f}\delta({\bf
x}-\mbox{\boldmath$\varepsilon$})$ with
$\mbox{\boldmath$\varepsilon$}=\varepsilon\unit{f}$, and then take
the limit $\varepsilon\rightarrow0$. However, this ignores the
boundary at radius~$R$, which should be kept fixed but is shifted
a distance $\varepsilon$ by the above procedure. An exact
calculation would require the monopole solution for force applied
near to (but not at) the center of a circular system, which, since
it no longer obeys radial symmetry, is likely to be highly
complex. Here we ignore such issues and simply use the above
monopole solution, in the  expectation that it will closely
approximate the exact case except possibly near the boundary. The
displacement  field induced by the force dipole is then
\begin{eqnarray}
u^{\rm dip}_{i}&=& -\frac{1}{\varepsilon} \mbox{\boldmath$\nabla$}
u^{\rm mono}_{i} \cdot{\mbox{\boldmath$\varepsilon$}}
\nonumber\\
&=& \frac{f}{8\pi\mu}\frac{1}{r}\bigg\{ \hat{r}_{i}
\left[4c_{1}-2c_{2}(r/R)^{-2}+2(r/R)^{2}\right]
\nonumber\\
&&\mbox{} +8\hat{r}_{i}({\unit{r}\cdot\unit{f}})^{2}
\left[c_{2}(r/R)^{-2}-c_{1}\right]
\nonumber\\
&&\mbox{} +2\hat{f}_{i}({\unit{r}\cdot\unit{f}})
\left[2(c_{1}-1)-2c_{2}(r/R)^{-2}\right.
\nonumber\\
&&\mbox{}\hspace{2cm} \left.-(c_{2}-1)(r/R)^{2}\right] \bigg\}\,.
\end{eqnarray}
As explained in Sec.~\ref{ss:decomposition}, the dipole solution
above applies for a single realization on large length scales, but
after averaging over many networks on short lengths (as in the
simulations) the dipole moment vanishes, leaving a quadrupole
displacement field. The required quadrupole is one consisting of
two parallel dipoles of equal and opposite magnitude, aligned
along their axes, {\em i.e.}
\begin{eqnarray}
u^{\rm quad}_{i}&=& \frac{1}{\varepsilon} \mbox{\boldmath$\nabla$}
u^{\rm dip}_{i} \cdot{\mbox{\boldmath$\varepsilon$}}
\nonumber\\
&=& \frac{f}{8\pi\mu}\frac{1}{r^{2}}\bigg\{ \hat{f}_{i}
\left[4(2c_{1}-1)-6c_{2}(r/R)^{-2}\right.
\nonumber\\
&&\mbox{}\hspace{2.5cm} +\left.2(2-c_{2})(r/R)^{2}\right]
\nonumber\\
&&\mbox{} +24\hat{r}_{i}({\unit{r}\cdot\unit{f}})
\left[c_{2}(r/R)^{-2}-c_{1}\right]
\nonumber\\
&&\mbox{} +8\hat{f}_{i}({\unit{r}\cdot\unit{f}})^{2}
\left[1-2c_{1}+3c_{2}(r/R)^{-2}\right]
\nonumber\\
&&\mbox{} +16\hat{r}_{i}({\unit{r}\cdot\unit{f}})^{3}
\left[2c_{1}-3c_{2}(r/R)^{-2}\right] \bigg\}\,.
\end{eqnarray}
This gives the displacement field in response to known forces of
magnitude~$f$. Since it is rather the displacement that is
controlled, $f$ is a free parameter. Finally, the non--zero modes
are
\begin{eqnarray}
h_{0}^{(r)}&=& \frac{\mu_{\rm affine}}{\pi\mu}\frac{1}{r^{2}}
\Big\{-c_{1}\Big\}\,,
\\
h_{2}^{(r)}&=& \frac{\mu_{\rm affine}}{2\pi\mu}\frac{1}{r^{2}}
\Big\{2c_{1}-3c_{2}(r/R)^{-2}\Big\}\,,
\\
h_{0}^{(f)}&=& \frac{\mu_{\rm affine}}{4\pi\mu}\frac{1}{r^{2}}
\Big\{3c_{2}(r/R)^{-2}
\nonumber\\
&&\mbox{}\hspace{2cm}+(2-c_{2})(r/R)^{2}\Big\}\,,
\\
h_{2}^{(f)}&=& \frac{\mu_{\rm affine}}{4\pi\mu}\frac{1}{r^{2}}
\Big\{1-2c_{1}+3c_{2}(r/R)^{-2}\Big\}\,.
\end{eqnarray}


\end{document}